\documentclass[prb,10pt,notitlepage,twocolumn]{revtex4}

\usepackage{graphicx}
\usepackage{amsmath}
\usepackage{amssymb}
\usepackage{amsfonts}
\usepackage{color}
\usepackage{xspace}
\usepackage{subfigure}
\usepackage{bm}
\usepackage{bbold}

\newcommand{\ket}[1]{| #1 \rangle}

\newcommand{\HgBrlong}{$\kappa$-(BEDT-TTF)$_2$Hg(SCN)$_2$Br\xspace}
\newcommand{\HgCllong}{$\kappa$-(BEDT-TTF)$_2$Hg(SCN)$_2$Cl\xspace}

\newcommand{\HgBr}{$\kappa$-HgBr\xspace}
\newcommand{\HgCl}{$\kappa$-HgCl\xspace}
\newcommand{\Hg}{$\kappa$-(BEDT-TTF)$_2$Hg(SCN)$_2Y$\xspace}
\newcommand{\X}{$\kappa$-(BEDT-TTF)$_2X$\xspace}
\newcommand{\CuCllong}{$\kappa$-(BEDT-TTF)$_2$Cu[N(CN)$_2$]Cl\xspace}

\newcommand{\CuCl}{$\kappa$-CuCl\xspace}
\newcommand{\CuBrlong}{$\kappa$-(BEDT-TTF)$_2$Cu[N(CN)$_2$]Br\xspace}

\newcommand{\CuBr}{$\kappa$-CuBr\xspace}
\newcommand{\CuCNlong}{$\kappa$-(BEDT-TTF)$_2$Cu$_2$(CN)$_3$\xspace}
\newcommand{\AgCNlong}{$\kappa$-(BEDT-TTF)$_2$Ag$_2$(CN)$_3$\xspace}

\newcommand{\CuCN}{$\kappa$-CuCN\xspace}

\newcommand{\AgCN}{$\kappa$-AgCN\xspace}

\renewcommand{\cite}[1]{[\onlinecite{#1}]}

\begin{document}

\title{Interplay of dipoles and spins in $\kappa$-(BEDT-TTF)$_2X$, where $X=$ Hg(SCN)$_2$Cl, Hg(SCN)$_2$Br, Cu[N(CN)$_2$]Cl, Cu[N(CN)$_2$]Br, and Ag$_2$(CN)$_3$
}
\author{A. C. Jacko}\affiliation{School of Mathematics and Physics, The University of Queensland, Brisbane, Queensland 4072, Australia}
\author{E. P. Kenny}\affiliation{School of Mathematics and Physics, The University of Queensland, Brisbane, Queensland 4072, Australia}
\author{B. J. Powell}\affiliation{School of Mathematics and Physics, The University of Queensland, Brisbane, Queensland 4072, Australia}

\begin{abstract}
We combine first principles density functional calculations with empirical  relationships for the Coulomb interactions in the  `monomer' model of $\kappa$-(BEDT-TTF)$_2X$. This enables us to calculate the parameters for the model of coupled dipolar and spin degrees of freedom  proposed by Hotta [Phys. Rev. B \textbf{82}, 241104 (2010)], and Naka and Ishihara [J. Phys. Soc. Japan \textbf{79}, 063707 (2010)]. In all materials studied, retaining only the largest  interactions leads to a transverse field Ising model of the dipoles. This quantifies, justifies and confirms recent claims that the dipoles are of crucial importance for understanding these materials.
We show that two  effects are responsible for a range of behaviors found in the dipoles  in different  \X salts. (i) The inter-dimer hopping, $t_{b1}$, which gives rise to the ``transverse field" in the Ising model for the dipoles ($H^T=2t_{b1}$), is between a third and a tenth smaller in the mercuric materials than for the mercury-free salts. (ii) The Ising model of dipoles is in the quasi-one-dimensional limit for the mercuric salts, but quasi-two-dimensional (between the square and isotropic triangular limits) for the mercury-free materials. Thus, the dimensionless critical field is much smaller in the mercuric compounds. Effect (ii) is  much larger than  effect (i). Simple explanations of both effects based on the band structures of the different salts are given.
We show that dipolar order and even short-range dipolar correlations have a profound impact on the nature of the interdimer magnetic (superexchange) interactions. For example,  dipole crystallization drives the materials towards quasi-one-dimensional magnetic interactions, which could be important for understanding the spin liquids found in some of these materials.
\end{abstract}

\maketitle

\section{Introduction}

%

\begin{figure*}
	\begin{center}
		\includegraphics[width=1.8\columnwidth]{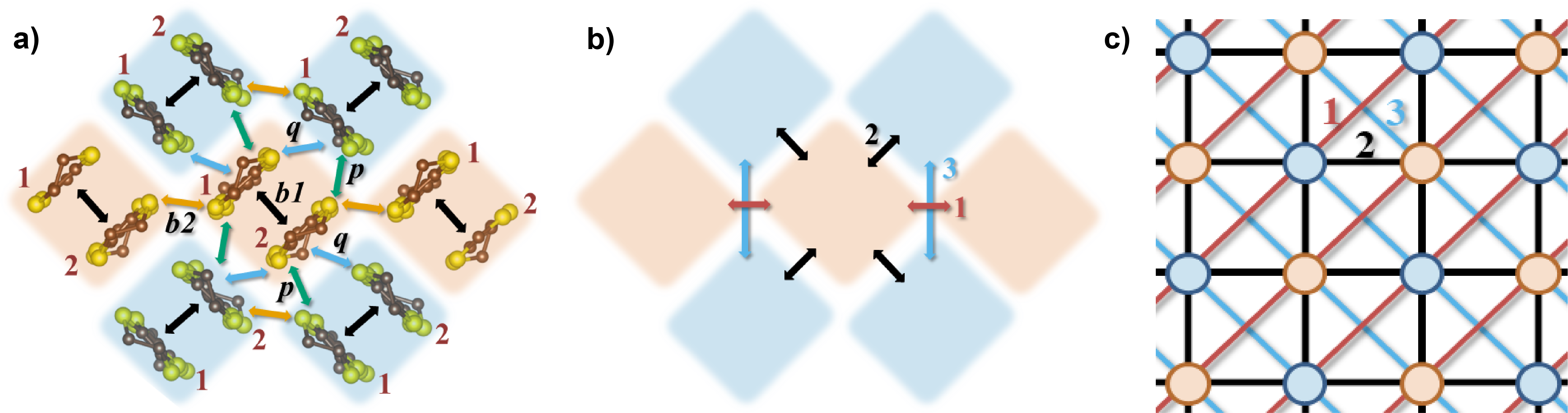} 
		\caption{Structure of models of \X. a) A single layer of \X  (hydrogen atoms and counter-ions are not shown). The various inter-molecular hopping integrals, \textit{i.e.}, $t_{\mu\nu ij}$ are labeled according to the standard monomer convention and, for simplicity, only the subcripts $\mu\nu ij\in\{b1, b2, p, q\}$ are shown. The two sublattices of ET dimers are highlighted in peach and blue. b) The dimer model. Here the subscripts 1, 2, or 3 will be used label various interactions, e.g., hopping $t$, exchange $J$, dipole-dipole coupling $K$, or Dzyaloshinskii-Moriya coupling $\bm D$. c) This model is topologically equivalent to a square lattice with different couplings across the two diagonals or, if the `3' interactions are negligible, an anisotropic triangular lattice.}\label{fig:lattice} 
	\end{center}
\end{figure*}

Which, if any, symmetries are spontaneously broken in the dimer Mott insulating phase of the \X depends crucially on the anion, $X$. \CuCllong (henceforth \CuCl) and \CuBrlong (henceforth \CuBr) show antiferromagnetic order, but no long-range magnetic order is observed in \CuCNlong (henceforth \CuCN) or \AgCNlong (henceforth \AgCN) down to the lowest temperatures that have been investigated. The standard model of these materials \cite{kino96,mckenzie98,powell06} begins from an (extended) Hubbard model with one orbital per BEDT-TTF, Fig. \ref{fig:lattice}a. This model is three quarters filled because the anion removes one electron per formula unit from the organic layer. However, as these salts are strongly dimerized, it is often assumed that the interdimer bonding and antibonding orbitals are energetically well separated. This assumption suggests that one can neglect the fully occupied bonding orbitals, leaving a half-filled Hubbard model with one orbital per dimer (BEDT-TTF)$_2$, Fig. \ref{fig:lattice}b,c.  

In the Mott insulating phase of the dimer model, charge fluctuations can be integrated out to leave a Heisenberg model on the anisotropic triangular lattice Fig. \ref{fig:lattice}c \cite{mckenzie98,powell11}. This model is  of intrinsic theoretical interest and it extrapolates between the square lattice ($J_1=0$), the triangular lattice ($J_1=J_2$) and one-dimensional chains with a frustrated interchain coupling ($J_2\rightarrow0$).

However, the validity of the dimer model has been questioned. Several groups have argued theoretically that the charge degrees of freedom within the dimer cannot be neglected and may  be responsible for some of the most interesting behavior observed in  \X, including the spin-liquid behavior and superconductivity \cite{Clay2019,Hotta,Naka}. Experimental support for these ideas has come from the observation of ferroelectricity arising from electronic dipoles associated with charge order within the dimers that spontaneously break inversion symmetry in several $\kappa$-(BEDT-TTF)$_2X$ salts \cite{Lunkenheimer,Sasaki,Abdel-Jawad}.

Recently, the \Hg salts have emerged as an important testing ground for the interplay between dipole and spin order. 
Long-range dipole order has been obeserved in \HgCllong (henceforth \HgCl) \cite{DrichkoPRB,Gati} and, collectively, several experiments suggest that a dipole liquid has been seen in \HgBrlong (henceforth \HgBr) \cite{DrichkoScience}.
Interestingly, it has been suggested that these materials may be an intermediate class with weaker dimerization than other $\kappa$-phase salts  \cite{DrichkoPRB,Gati}, but stronger dimerization than, say, the $\theta$ phase BEDT-TTF salts \cite{SeoCR}. 

One approach to understanding the interaction between spin and charge order in the \X salts is to study the monomer model (Fig. \ref{fig:lattice}) \cite{Clay2019}. But an alternative,  physically appealing, approach has been introduced by Hotta \cite{Hotta} and, independently, Naka and Ishihara \cite{Naka} (HIN). They introduce a  binary variable, which describes the dipole or charge degree of freedom within the dimer. HIN showed that this leads to a transverse field Ising model (TFIM) for the dipoles, a Heisenberg model for the spin on the dimers, and coupling terms between the spin and dipole degrees of freedom, reminiscent of the Kugel-Khomskii interaction between spins and orbital degrees of freedom \cite{Pavani}. 

In this paper we derive a first principles tight-binding monomer model for \HgCl, \HgBr, \CuCl, \CuBr, and \AgCN. We use this model, and empirical relationships for the effective Coulomb interactions, to parameterise an extended HIN model for these materials. We find that, in all materials, the coupling constants for the TFIM of the dipoles are the largest energy scales in the problem.

The TFIM has been extensively studied in each of the limits of the lattice dimer model, Fig. \ref{fig:lattice}. In one dimension, the dimensionless critical field is $h_c\equiv H^T_c/K_\text{max}=1$, where $H^T_c$ is the critical transverse field and $K_\text{max}$ is the (largest) Ising coupling constant. On the isotropic triangular lattice $h_c=4.768$ \cite{squareIsing02} and on the square lattice $h_c=3.044$  \cite{squareIsing98,squareIsing02}. However, we are not aware of any studies on the anisotropic triangular lattice, Fig. \ref{fig:lattice}c, beyond these limits. 
In the TFIM description of the dipoles in \X, there is a direct proportionality between the transverse field and the intradimer hopping ($H^T\propto t_{b1}$). There is ferroelectric order in the low field regime and quantum disorder in the high field regime.

We show that two  effects are responsible for different ferroelectric behaviors of the \X salts. (i) The inter-dimer hopping, $t_{b1}$, which gives rise to the ``transverse field" in the Ising model for the dipoles ($H=2t_{b1}$), is between a third and a tenth smaller in  \HgCl and \HgBr than in \CuCl, \CuBr, and \AgCN. (ii) The TFIM of dipoles is in the quasi-one-dimensional limit for \HgCl and \HgBr, but quasi-two-dimensional (between the square and isotropic triangular limits) for \CuCl, \CuBr, and \AgCN. Thus, the dimensionless critical field is much smaller in the former pair of compounds.  We discuss the implications of this for the magnetic behaviors of these salts.

\section{Density functional calculations}

We computed the electronic structures of \HgCl, \HgBr, \CuCl, \CuBr, and \AgCN from four-component relativistic density functional theory with the FPLO package  in an all-electron full-potential local orbital basis \cite{eschrig04,koepernik99}. For \CuBr we consider fully dueterated BEDT-TTF molecules (as this drives this salt insulating \cite{taniguchi99}), for all other anions we consider fully protonated BEDT-TTF. The density was converged on a $(8 \times 8 \times 8)$ $k$ mesh using the Perdew, Burke and Ernzerhof (PBE) generalized gradient approximation \cite{perdew96}. 
We produced localised Wannier spin-orbitals for the frontier bands (those closest to the Fermi energy), and from those produced an \textit{ab initio} single electron Hamiltonian containing one Wannier orbital, $|\mu i\alpha\rangle$, per spin species, $\alpha$, on every BEDT-TTF molecule, $i$ in every dimer, $\mu$. 
The geometries were taken from the measured crystal structures \cite{DrichkoPRB,SchlueterPrivate,Geiser91,GeiserActa91,SaitoPrivate} as indicated in Table \ref{tab:ts}.


The complex overlaps between the Wannier spin-orbitals, 
\begin{eqnarray}
h_{\mu\nu ij\alpha\beta}=\langle \mu i\alpha|\hat{\mathcal H}|\nu j\beta\rangle,
\end{eqnarray} 
where $\hat{\mathcal H}$ is the full microscopic Hamiltonian,  lead to a spin-dependent model Hamiltonian:
\begin{equation}
\hat{H} = \sum_{\mu,\nu} \sum_{i,j} \sum_{\alpha,\beta}\hat c^\dagger_{\mu i\alpha} h_{\mu\nu ij\alpha\beta}\hat c_{\nu j\beta}, 
\end{equation}
where $\hat c^{(\dagger)}_{\mu i\alpha}$ annihilates (creates) an electron in the spin $\alpha$ Wannier centered on the $i$th molecule of the $\mu$th dimer. The SU(2) symmetry of the Hamiltonian implies that $h_{\mu\nu ij} = t_{\mu\nu ij}I_2 + \bm{\lambda}_{\mu\nu ij}\cdot\bm{\sigma}$, where $I_2$ is the identity matrix and $\bm{\sigma}$ is the vector of Pauli matrices. Thus,
\begin{equation}
\hat{H} = \sum_{\mu,\nu} \sum_{i,j} \sum_{\alpha,\beta}\hat  c^\dagger_{\mu i\alpha} \Big(t_{\mu\nu ij}  \delta_{\alpha \beta} + i \bm{\lambda}_{\mu\nu ij}\cdot\bm{\sigma}_{\alpha \beta} \Big)\hat  c_{\nu j\beta}. \label{eq:Hrel}
\end{equation}
Physically, the scalar hopping integrals, $t_{\mu\nu ij}$, describe  spin independent hopping between molecules and the vector hopping integrals,  $\bm{\lambda}_{\mu\nu ij}$, which parameterize the spin-orbit coupling, allow for both spin dependent hopping ($\lambda^z_{\mu\nu ij}$) and hopping with a spin flip ($\lambda^x_{\mu\nu ij}$, $\lambda^y_{\mu\nu ij}$).

\begin{figure}
	\begin{center}
		\includegraphics[width=0.9\columnwidth]{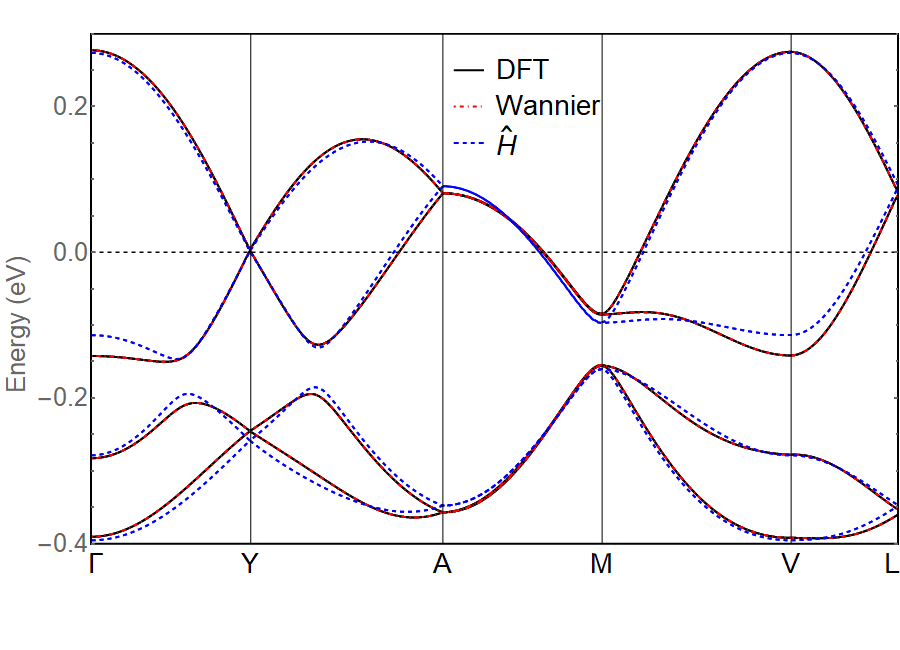} 
		\caption{The calculated band structure of \HgCl. The direct density functional calculations (solid black line) is exactly reproduced by the Wannier model Hamiltonian (dot-dashed red line). The monomer model [$\hat H$, blue dashed line, Fig. \ref{fig:lattice}, Eq. (\ref{eq:Hrel})], which retains only the four largest hopping integrals, captures the main features of the band structure.  }
		\label{fig:BS-Cl}
	\end{center}
\end{figure}

A typical band structure is shown in Fig. \ref{fig:BS-Cl}. We
display three different approximations to the electronic structure: (i) the band structure directly calculated from four component DFT; (ii) the band structure calculated from the full set of overlaps between  Wannier orbitals, $\{h_{\mu\nu ij\alpha\beta}\}$; and (iii) the standard monomer model, Fig. \ref{fig:lattice}a, which contains only the largest overlaps of our \textit{ab initio} model Hamiltonian. As expected in both materials the full Wannier band structure  reproduces the direct DFT results exactly, consistent with a successful Wannier orbital construction. The monomer model clearly captures the key features of the electronic structure and gives a clearer picture of the physical mechanisms at play than the full DFT calculation. Therefore, we will use this model as the basis for analysis of these materials.
Also, note the Dirac points at $Y$ in \HgCl and \HgBr (not shown).


%

The monomer model is based on Eq. (\ref{eq:Hrel}) but restricted to contain only the largest parameters ($t_{\mu\nu ij}$ and $\bm{\lambda}_{\mu\nu ij}$). As in other $\kappa$-phase BEDT-TTF salts, there are four large interactions in both of the \Hg salts. We  use the standard bond-labeling convention, illustrated in Fig. \ref{fig:lattice}a. The parameters for this model are reported in Table \ref{tab:ts}.  The scalar hopping integrals for these salts are in excellent agreement with a previous scalar relativistic and non-relativistic calculations where these have previously been reported \cite{Gati,Hotta-params}.

\begin{table*}
	\centering
\begin{tabular}{l|ccccccccccc}   
				& $t_{b1}$ & $t_{b2}$ 	& $t_p$ &$ t_q$	& $\sqrt{t_p^2-t_q^2}$ & 
${\bm\lambda}_p$ &$ {\bm\lambda}_q$ & $\bm{r}_{p}$	& $\bm{r}_{q}$ 	\\\hline
\HgCl \cite{DrichkoPRB} 		& 120		& 88		& 63 	& 34	& 33 	
& $(0.83, -0.07, 0.32)$ 	& $(0.37, -0.13, 0.08)$ & $(0.00, -2.72, 11.09)$  & $(3.60, -7.67, -7.59)$  	 \\
\HgBr \cite{SchlueterPrivate} 		& 126 		& 83		& 60	& 40	& 45	
	& $(0.40, 0.06, 0)$ 	& $(1.47, 0.10, -0.53)$ & $(0.00, -2.76, 11.01)$  & $(3.47, -7.75, 7.51)$  \\
\CuCl \cite{Geiser91}	& 142 		& 60 		& 90 	& 44	& 79	
	& $(-0.54, -0.91, 0)$ 	& $(0.37, -0.68, 0)$ & $(7.61, -0.99, 7.95)$  & $(-12.12, 0.00, 3.47)$  \\
\CuBr  \cite{GeiserActa91}	& 178 		& 61 		& 98 	& 37	& 91	
	& $(-0.46, -0.87, 0)$ 	& $(0.30, -0.84, 0)$ &  $ (7.58, -0.99, 8.00)$ & $(-12.16, 0.00, 3.53)$   \\
\AgCN \cite{SaitoPrivate}		& 180 		& 66 		& 87 	& 19	& 85	
	& $(-0.55, -0.01, 0.98)$ 	& $(-0.03, -0.29, -0.07)$  & $(-1.03, 8.16, -7.84)$  &  $(0.00, 3.70, 12.43)$  \\
\end{tabular}
	\caption{Calculated intra-layer scalar ($t$) and vector ($\bm\lambda$) hopping integrals in meV and the relevant vectors in \AA. ${\bm\lambda}_{b1}={\bm\lambda}_{b2}=\bm0$ in all materials due to inversion symmetries.  $\bm{r}_{p}$ and $\bm{r}_{q}$ define the direction of the vector hopping [reversing the direction changes the sign -- as required for the hermiticity of Eq. (\ref{eq:Hrel})].  $\sqrt{t_p^2-t_q^2}$ is included here as, we will see below,  it is a crucial parameter for understanding the behavior of these materials. Note that it is much smaller in \HgCl and \HgBr than in the other materials.}\label{tab:ts}
\end{table*}

\section{Coupled spin and dipole order}

 In this section we  generalize the model of interacting dipoles and spins introduced by Hotta, Naka and Ishihara \cite{Naka,Hotta} and use our density functional theory results as the basis for a calculation of the parameters of the model.
The derivation assumes that these materials are in the strong coupling limit where the electron-electron interactions within the dimer are large, and all other terms can be treated perturbatively. The unperturbed model is
\begin{eqnarray}
\hat{H}^{(0)}&=&\sum_\mu \Bigg[
 U \sum_i \hat n_{\mu i\uparrow} \hat n_{\mu i\downarrow}
+ \left(V_{b1}-\frac{J_{b1}}{2}\right) \hat n_{\mu 1} \hat n_{\mu 2} \notag \\ && \notag
 + J_{b1}\left( \hat{c}_{\mu1\uparrow}^\dagger \hat{c}_{\mu1\downarrow}^\dagger  \hat{c}_{\mu2\uparrow} \hat{c}_{\mu2\downarrow} 
 				+ \hat{c}_{\mu2\uparrow}^\dagger \hat{c}_{\mu2\downarrow}^\dagger  \hat{c}_{\mu1\uparrow} \hat{c}_{\mu1\downarrow} \right)\\ &&
 				- 2J_{b1} \hat{\bm S}_{\mu 1} \cdot \hat{\bm S}_{\mu 2}\Bigg] , 
\end{eqnarray} 
where $i=1$ and $2$ indicate the two BEDT-TTF molecules within the same dimer, $\hat n_{\mu i\alpha}=\hat c^\dagger_{\mu i\alpha} \hat c_{\mu i\alpha}$, $\hat n_{\mu i} = \sum_{\alpha} \hat n_{\mu i\alpha}$, and  $\hat{\bm S}_{\mu i} = \frac12 \sum_{\alpha,\beta} \hat c^\dagger_{\mu i\alpha} \bm{\sigma}_{\alpha\beta} \hat c_{\mu i\beta}$.
At three-quarters filling, relevant to  \X, there are four degenerate ground states: $\ket{\uparrow\downarrow,\uparrow}$,  $\ket{\uparrow\downarrow,\downarrow}$,  $\ket{\uparrow,\uparrow\downarrow}$, and  $\ket{\downarrow,\uparrow\downarrow}$.

Intra-dimer hopping and inter-dimer interactions  give rise to first order perturbations. We write their sum as
\begin{eqnarray}
\hat{H}^{(1)} &=& t_{b1} \sum_{\mu\sigma} \hat c^\dagger_{\mu 1\sigma} \hat c_{\mu 2\sigma} \\\notag &&
+ \frac12 \sum_{\mu\ne\nu} \sum_{i,j} \Bigg[ 
 \left( V_{\mu\nu}^{ij} - \frac{J_{\mu\nu}^{ij}}{2} \right) \hat n_{\mu i} \hat n_{\nu j}
\notag \\ \notag
&& - 2J_{\mu\nu}^{ij} \hat{\bm S}_{\mu i} \cdot \hat{\bm S}_{\nu j} 
+ J_{\mu\nu}^{ij} \hat{c}_{\mu i\uparrow}^\dagger \hat{c}_{\mu i\downarrow}^\dagger  \hat{c}_{\nu j\uparrow} \hat{c}_{\nu j\downarrow} 
\Bigg].
\end{eqnarray}
Throughout we adopt the convention that within any pair of dimers the labels $i$ and $j$ are chosen so that the shortest separation connects site 1 on one dimer with site 2 on the other, this preserves translation symmetry within each sublattice, cf. Fig. \ref{fig:lattice}.

Within the low-energy subspace these terms can be dramatically simplified by the introduction of two new operators: the dimer spin operator, $\hat{\bm S}_{\mu} = \frac12 \sum_{i,\alpha,\beta} \hat c^\dagger_{\mu i\alpha} \bm{\sigma}_{\alpha\beta} \hat c_{\mu i\beta}$,  and the dipole operator $\hat{\bm T}_{\mu} = \frac12 \sum_{i,j,\alpha} \hat c^\dagger_{\mu i\alpha} \bm{\sigma}_{ij} \hat c_{\mu j\alpha}$. Our four degenerate ground states can be rewritten as $(S^z,T^z)=(\frac12,\frac12)$, $(-\frac12,\frac12)$, $(\frac12,-\frac12)$, and $(-\frac12,-\frac12)$ respectively. This allows us to write
\begin{eqnarray}
\hat{H}^{(1)} &=& H^T \sum_\mu \hat{T}^x_\mu 
+ \sum_{\mu\ne\nu}  W^{TT}_{\mu\nu} \hat{T}^z_\mu\hat{T}^z_\nu  \notag\\ &&
+ \sum_{\mu\ne\nu} \left[ W_{\mu\nu}^{SS} 
+ W_{\mu\nu}^{SST}(\hat{T}^z_\mu + \hat{T}^z_\nu) \right.\notag\\&&\left.\hspace{20mm}
+ W_{\mu\nu}^{SSTT} \hat{T}^z_\mu \hat{T}^z_\nu \right] \hat{\bm S}_{\mu} \cdot \hat{\bm S}_{\nu}, \label{eq:H1order}
\end{eqnarray}
where 
$H^T=2t_{b1}$,
\begin{eqnarray}
W^{TT}_{\mu\nu} &=& ( V_{\mu\nu}^{11} - V_{\mu\nu}^{12} - V_{\mu\nu}^{21} + V_{\mu\nu}^{22} ) \notag\\&& -\frac12 ( J_{\mu\nu}^{11} - J_{\mu\nu}^{12} - J_{\mu\nu}^{21} + J_{\mu\nu}^{22} ), \label{eq:WTT}
\end{eqnarray}
$W_{\mu\nu}^{SS} = \frac14( J_{\mu\nu}^{11} + J_{\mu\nu}^{12} + J_{\mu\nu}^{21} + J_{\mu\nu}^{22} )$, 
$W_{\mu\nu}^{SST} = \frac12( J_{\mu\nu}^{11} + J_{\mu\nu}^{12} - J_{\mu\nu}^{21} - J_{\mu\nu}^{22} )$,
$W_{\mu\nu}^{SSTT} =  J_{\mu\nu}^{11} - J_{\mu\nu}^{12} - J_{\mu\nu}^{21} + J_{\mu\nu}^{22}$,
and we have used the symmetries $V_{\mu\nu}^{ij}=V_{\nu\mu}^{ji}$, $J_{\mu\nu}^{ij}=J_{\nu\mu}^{ji}$ and the inversion symmetry of the crystals (which, although  spontaneously broken in the dipoled ordered phase, is respected by the underlying Hamiltonian). 

The first two terms of Eq. (\ref{eq:H1order}) are simply the TFIM for the dipole variables, $\hat{\bm T}_\mu$. The last two lines describe direct dimer exchange interactions  ($W_{\mu\nu}^{SS}$) and three ($W_{\mu\nu}^{SST}$) and four ($W_{\mu\nu}^{SSTT}$) body interactions between the dipole Ising variables and the spins. The interdimer exchange interactions ($J_{\mu\nu}^{ij}$) are around two orders of magnitude smaller than direct interdimer Coulomb interactions  ($V_{\mu\nu}^{ij}$), so the first two terms in Eq. (\ref{eq:H1order}) are significantly larger than others (although we see below that these terms are not negligible for describing the magnetic order). Therefore it is reasonable to start by examining only the TFIM for $\hat{T}^x_\mu$. It is also important to note that the Ising coupling constants, $W_{\mu\nu}^{TT}$, are given by a ``differential'' Coulomb interaction, Eq. (\ref{eq:WTT}).

It has been demonstrated that in full three-dimensional models of BEDT-TTF salts $V_{\mu\nu}^{ij}\propto1/r$, to a high degree of accuracy \cite{Shinaoka}, where $r$ is the distance between the centers of the Wannier orbitals, and $V_{b2}\sim0.4$ in a wide range of salts \cite{Hotta-params,Nakamura09}. These two relationships yield the Ising coupling constants listed in Table \ref{tab:Ising}. 
The differences between the different materials result primarily from changes in the way the dimers are arranged within the crystal. These structural differences are much larger than changes of the intradimer separation, which only varies between 3.67 and 3.73 \AA\xspace across all materials studied.

It is apparent from  Eq. (\ref{eq:WTT}) that the $W^{TT}$ can take either sign. However, one's physical intuition is that if the dimers are close to each other (relative to the intradimer separation) then $W^{TT}$ should be negative. This is indeed confirmed by the values in Table \ref{tab:Ising}, with the only positive values being $W_3^{TT}$ for \HgCl and \HgBr and $W_\perp^{TT}$ for \AgCN. In each of these cases the interdimer separation is large compared to intradimer separation and the argument above becomes invalid. Given the signs in the first order Hamiltonian, Eq. (\ref{eq:H1order}), (positive) negative $W^{TT}$ corresponds to (anti)ferro-dipolar coupling. Nevertheless, because of the relative arrangement of the  sublattices of the dimers, cf. Fig. \ref{fig:lattice}a, long range (anti)ferro-dipolar order does not correspond to simple (anti)parallel arrangements of the dipoles on the two different sublattices.

\begin{table}
	\centering
	\begin{tabular}{l|cccc}   
		$X$  & $W^{TT}_{1}$ & $W^{TT}_{2}$ 	& $W^{TT}_3$ & $W^{TT}_{\perp}$ 		\\ \hline
		Hg(SCN)$_2$Cl   & -8 	& -22 	& 13  	& -3 		\\
		Hg(SCN)$_2$Br  & -11 	& -21 	& 13 	& -3 		\\
		Cu[N(CN)$_2$]Cl	& -5 	& -54 	& -12 	& -11 	\\
		Cu[N(CN)$_2$]Br	& -5 	& -54  	& -12 	& -11 	\\
		Ag$_2$(CN)$_3$	& -4 	& -53 	& -12 	& 1 
	\end{tabular}
	\caption{Calculated Ising coupling constants at first order in the interdimer Coulomb interactions, cf. Eq. (\ref{eq:WTT}), for the dipole degrees of freedom. We neglect the $J_{\mu\nu}^{ij}$  as these are two orders of magnitude smaller than the $V_{\mu\nu}^{ij}$ \cite{nakamura12}. All values in meV.}\label{tab:Ising} 
\end{table}

Virtual hopping between dimers induces additional interactions. Hopping within a single dimer sublattice, cf. Fig. \ref{fig:lattice}, is described by 
\begin{eqnarray}
	\hat{H}_1 = t_{b2}\sum_{\langle\mu,\nu\rangle}\sum_{\sigma} \hat c^\dagger_{\mu 2\sigma} \hat c_{\nu 1\sigma} + H.c., \label{eq:tb2}
\end{eqnarray}
where $\langle\mu,\nu\rangle$ indicates that the sum runs over nearest neighbors on the same sublattice only.  At second order in $t_{b2}$ this induces  interactions given by
\begin{eqnarray}
\hat{H}^{(2)}_{1} &=&  
\sum_{\langle\mu,\nu\rangle} \Big\{ 
X_{TT} \hat{T}^z_\mu\hat{T}^z_\nu 
+  X_{T} (\hat{T}^z_\mu - \hat{T}^z_\nu)
 \notag\\ &&  \hspace{7mm}
+ \Big[  X_{SSTT} \hat{T}^z_\mu\hat{T}^z_\nu 
+ X_{SST} (\hat{T}^z_\mu - \hat{T}^z_\nu) 
 \notag\\ &&   \hspace{12mm}
+ X_{SS} 
\Big] \hat{\bm S}_{\mu} \cdot \hat{\bm S}_{\nu} \Big\}. \label{eq:X}
\end{eqnarray}
Analytic expressions for the coupling constants are given in the Appendix.

\begin{figure}
	\begin{center}
		\includegraphics[width=0.9\columnwidth]{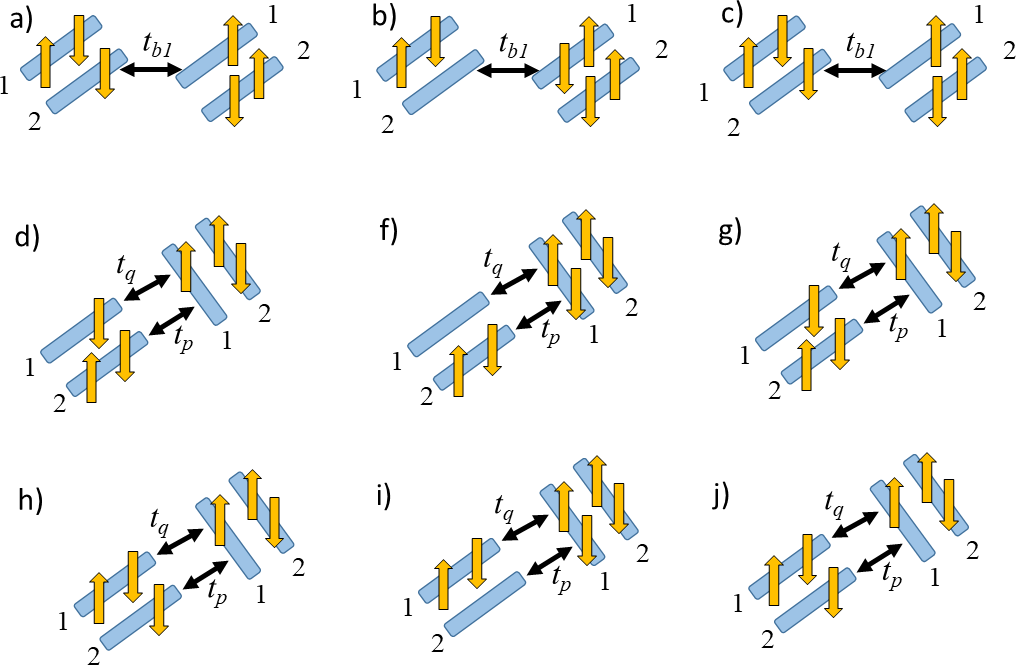}
		\caption{Sketches of prototypical second order processes leading to the Ising coupling between dipoles. Recall that second order processes always lower the energy of the ground state. Processes like $a\rightarrow b\rightarrow c$ are only available if dipoles on the two dimes are antiparallel. Thus, this leads to an antiferro-dipolar coupling ($X_{TT}<0$).  Processes like $d\rightarrow e\rightarrow f$ are only allowed if dipoles on the two dimes are  the same sign and therefore favor ferro-dipolar coupling ($Y_{TT}>0$). However, processes like $h\rightarrow i\rightarrow j$ are only possible if dipoles on the two dimes are antialigned and therefore favor antiferro-dipolar coupling ($Y_{TT}<0$). The competition between the processes $d\rightarrow e\rightarrow f$ and $h\rightarrow i\rightarrow j$ is vital for understanding the different dimensionalities of the TFIM in the mercuric and mercury-free salts as it means the $Y_{TT}\propto t_p^2-t_q^2$.}\label{fig:whyXTT} 
	\end{center}
\end{figure}

A key observation from these analytic expressions is that $X_{TT}$ is generically positive, i.e., antiferro-dipolar. This is the opposite sign from that expected (and calculated) for $W_1^{TT}$, the equivalent interaction arising  from the direct Coulomb interaction. However, it is the same sign as one expects from a simple superexchange interaction, $X_{SS}$, which also arises at second order in $t_{b2}^2/U$. The processes that give rise to this $X_{TT}$ are sketched in Fig. \ref{fig:whyXTT}.

Hopping between different dimer sublattices is described by the terms
\begin{eqnarray}
\hat{H}_2 = \sum_{[\mu,\nu]}\sum_{\sigma} \left( t_{p} \hat c^\dagger_{\mu 2\sigma} \hat c_{\nu 1\sigma} + t_{q} \hat c^\dagger_{\mu 2\sigma} \hat c_{\nu 2\sigma} \right) + H.c.,  \label{eq:tpq}
\end{eqnarray}
where $[\mu,\nu]$ indicates that the sum runs over nearest neighbors on the different sublattices only. 

At second order in $t_p$ and $t_q$, the effective Hamiltonian is 
\begin{widetext}
\begin{eqnarray}
\hat{H}^{(2)}_{2} &=&  
\sum_{[\mu,\nu]} \Big\{ 
Y_{TT} \hat{T}^z_\mu\hat{T}^z_\nu 
+ Y_{zx} \hat{T}^z_\mu\hat{T}^x_\nu
+  Y_{T-} (\hat{T}^z_\mu - \hat{T}^z_\nu)
+  Y_{T+} (\hat{T}^z_\mu + \hat{T}^z_\nu)
+  Y_{x} \hat{T}^x_\nu
\notag\\ &&  
+ \Big[  Y_{SSTT} \hat{T}^z_\mu\hat{T}^z_\nu 
+ Y_{SSzx} \hat{T}^z_\mu\hat{T}^x_\nu 
+ Y_{SSx} \hat{T}^x_\nu
+ Y_{SST-} (\hat{T}^z_\mu - \hat{T}^z_\nu) 
+ Y_{SST+} (\hat{T}^z_\mu + \hat{T}^z_\nu) 
+ Y_{SS} \Big] \hat{\bm S}_{\mu} \cdot \hat{\bm S}_{\nu}
\notag\\ &&
+\Big[
Y_{4Szx} \hat{T}^z_\mu\hat{T}^x_\nu 
+ Y_{4Sx} \hat{T}^x_\nu
\Big] \left(\hat{\bm S}_{\mu} \cdot \hat{\bm S}_{\nu}\right)^2 \Big\}. \label{eq:Y}
\end{eqnarray}
\end{widetext}
Again, the analytical expressions for the coupling constants are given in the Appendix.

It is interesting to note here that the values of the $Y$ parameters, most importantly $Y_{TT}\propto (t_p^2 - t_q^2)$ [see Eq. (\ref{eq:YTT})], depend sensitively on the relative magnitudes of $t_p$ and $t_q$ (reported in Table \ref{tab:ts}). This has a simple physical origin, explained in Fig. \ref{fig:whyXTT}. This will be crucial for understanding the different physics observed in the different materials.

To evaluate the $X$ and $Y$ parameters we take $U=V_{b1}$, $J_{b1}=V_{b1}/2.5$    \cite{Nakamura09} and $V_{b1}$ given by the parameterization above \footnote{$V_{b1}=672$ meV for \HgCl, 677 meV for \HgBr, 685 meV for \CuCl, 683 meV for \CuBr, and 682 meV for \AgCN}. This yields the parameters in Table \ref{tab:Ising_pert}. Notice that the largest parameters here are  an order of magnitude more than those arising from the first order corrections, Table \ref{tab:Ising}. We stress that this does not indicate an issue with the convergence of the perturbative expansion, as these corrections are second order in inter-dimer hopping integrals and the parameters in Table \ref{tab:Ising} are first order in the inter-dimer Coulomb interactions.  

The two largest parameters, $X_{TT}$ and $Y_{TT}$, are quite different in \HgCl and \HgBr than in the other materials. $X_{TT}\propto t_{b2}^2$ [see Eq. (\ref{eq:XTT})] and $t_{b2}$ is 25-50\% larger in \HgCl and \HgBr than the other materials, see Table \ref{tab:ts}, which means that $X_{TT}$ is much bigger in the mercuric materials. 
$Y_{TT}\propto (t_p^2 - t_q^2)$ [see Eq. (\ref{eq:YTT})] and $t_p$ is $\sim50\%$ larger in the mercury-free materials, which results in a much larger $Y_{TT}$ in these systems.

\begin{table*}
	\centering
	\begin{tabular}{l|cccccccccccccccccc}   
		$X$  & $X_{TT}$ &  $X_{T}$ &  $X_{SSTT}$ &  $X_{SST}$ &  $X_{SS}$ & $Y_{TT}$ &  $Y_{zx}$ &  $Y_{Tm}$ &  $Y_{Tp}$ &  $Y_{x}$ &  $Y_{SSTT}$ &  $Y_{SSzx}$ &  $Y_{SST-}$ &  $Y_{SST+}$ &  $Y_{SSx}$ &  	$Y_{SS}$ &  $Y_{4Szx}$ &  $Y_{4Szx}$ 		\\ \hline
		Hg(SCN)$_2$Cl   & 288 & -6 & -46 & -23 & 23 & 105 & -40 & -3 & -1 & -20 & -17 & -38 & -12 & -3 & -19 & 8 & 26 & 26		\\
		Hg(SCN)$_2$Br   & 254 & -5 & -41 & -20 & 20 & 74 & -44 & -3 & -1 & -22 & -12 & -43 & -11 & -5 & -21 & 6 & 28 & 28	\\
		Cu[N(CN)$_2$]Cl	& 131 & -3 & -21 & -11 & 11 & 225 & -72 & -6 & -1 & -36 & -36 & -69 & -24 & -6 & -35 & 18 & 46 & 46 \\
		Cu[N(CN)$_2$]Br	& 136 & -3 & -22 & -11 & 11 & 301 & -66 & -7 & -1 & -33 & -48 & -64 & -28 & -4 & -32 & 24 & 42 & 42 \\
		Ag$_2$(CN)$_3$	&160 & -3 & -26 & -13 & 13 & 264 & -30 & -6 & 0 & -15 & -42 & -29 & -22 & -1 & -15 & 21 & 19 & 19 \\
	\end{tabular}
	\caption{Calculated coupling constants for the effective Hamiltonians, Eqs. (\ref{eq:X}) and (\ref{eq:Y}), arising at second order in the interdimer hopping. All values in meV.  See Appendix for the analytic expressions.}\label{tab:Ising_pert}
\end{table*}

Combining the lowest order perturbations in both the Coulomb interactions and the hopping but
keeping only the largest terms  reduces the problem to an effective TFIM:
\begin{eqnarray}
H_{TFI} = H^T \sum_\mu \hat{T}^x_\mu 
+ \sum_{\mu\ne\nu}  K_{\mu\nu} \hat{T}^z_\mu\hat{T}^z_\nu
\label{eq:HIsing}
\end{eqnarray}
with two large ($>100$ meV) coupling constants $K_1=W_1^{TT}+X_{TT}$ and $K_2=W_2^{TT}+Y_{TT}$. The values of these parameters are reported in Table \ref{tab:Ising_final}.

\begin{table}
	\centering
	\begin{tabular}{l|cccc}   
		$X$  			& $K_1$ [meV] &  $K_2$ [meV] & $K_1/K_2$	& $H^T/K_\text{max}$	\\ \hline
		Hg(SCN)$_2$Cl   & 280 & 83 & 3.38 & 0.86  \\
		Hg(SCN)$_2$Br   & 244 & 53 & 4.6 & 1.03  \\
		Cu[N(CN)$_2$]Cl	& 126 & 171 & 0.74 & 1.66 \\
		Cu[N(CN)$_2$]Br	& 131 & 248 & 0.53 & 1.44 \\
		Ag$_2$(CN)$_3$	& 155 & 211 & 0.73 & 1.70
	\end{tabular}
	\caption{The two largest coupling constants in the Ising model of dipoles, $K_1$ and $K_2$, and the ratio of the larger of these ($K_\text{max}$) with the effective transverse field, $H^T$. Recall that in one dimension the critical field for the transverse field Ising model is $h_c\equiv H_c^T/K=1$, in the isotropic triangular lattice $h_c=4.768$ \cite{squareIsing02} and on the square lattice $h_c=3.044$ \cite{squareIsing98,squareIsing02}.}\label{tab:Ising_final}
\end{table}

In both \HgCl and \HgBr $K_1\gg K_2$, placing these materials firmly in the quasi-one-dimensional limit of the TFIM. In \HgBr we find that $H^T/K_1>1$, \emph{i.e.}, the high-field regime of the TFIM, which corresponds to quantum disorder of the dipoles. But \HgBr is close to the phase transition. Therefore our calculations predict that there is no dipole order, but strong correlations between the dipoles. This is entirely consistent with the observation of a `dipole liquid' in this material \cite{DrichkoScience}.  

In \HgCl we find that $H^T/K_1<1$, \emph{i.e.}, the low-field regime of the TFIM, which corresponds to spontaneous ordering of the dipoles in the underlying extended Hubbard model. This explains the observed dipole solid in \HgCl \cite{DrichkoPRB}.

In \CuCl, \CuBr and \AgCN are intermediate between the square and triangular lattice TFIMs, $0.53\leq K_1/K_2\leq 0.74$. We are not aware of any studies of the TFIM in this regime, which makes it impossible to make definitive predictions of the experimental consequences of our results.  \CuCl and \AgCN yield rather similar TFIMs with strong frustration $K_1/K_2\sim3/4$, but the frustration is somewhat less in \CuBr. In all three materials $H^T/K_2$ is well below the critical fields of either the square or triangular TFIMs, therefore it is tempting to conclude that all three materials should show dipole order. But caution is needed here as we do not know how frustration effects the critical field. For example, in the Heisenberg model long-range order is strongly suppressed at frustrations  intermediate to the square and triangular limits \cite{Zheng1999,scriven12}.

\section{Spin model in the dipole solid phase}

To calculate the inter-dimer superexchange interactions in the dipole ordered phase we make a mean-field  approximation for the charge-charge interdimer interactions,
\begin{eqnarray}
\hat V_{\mu\nu} = \sum_{\mu\ne\nu} \sum_{i,j}  V_{\mu\nu}^{ij}  \hat n_{\mu i} \hat n_{\mu j},
\end{eqnarray}
which induces an effective potential on the electrons within the $\mu$th dimer given by
\begin{eqnarray}
 \hat{V}_{\mu\nu}^\text{MF} &\equiv& \left({\sum_{\nu}}' \sum_{i,j} V_{\mu\nu}^{ij}  \langle\hat n_{\nu j}\rangle \right)\hat n_{\mu i} \notag \\
	&=& \varepsilon^*\left(\hat n_1 + \hat n_2\right) - \Delta \left(\hat n_1 - \hat n_2\right),
\end{eqnarray}
where the prime on the first summation indicates that it runs over all dimers except $\mu$, 
\begin{eqnarray}
\varepsilon^* =    {\sum_{\nu }}'\sum_{i,j} \frac{V_{1\nu}^{ij} + V_{2\nu}^{ij}}{2} \langle\hat n_{\nu j}\rangle,
\end{eqnarray}
and the difference in potentials within a dimer is
\begin{eqnarray}
\Delta =    {\sum_{\nu }}'\sum_{i,j}  \frac{V_{1\nu}^{ij} - V_{2\nu}^{ij}}{2} \langle\hat n_{\nu j}\rangle.
\end{eqnarray}

Thus, the effective Hamiltonian for a single dimer is
\begin{eqnarray}
	\hat H_\mu &=& (\varepsilon_0 + \varepsilon^*)\left(\hat n_{\mu1} + \hat n_{\mu2}\right) - \Delta \left(\hat n_{\mu1} - \hat n_{\mu2}\right) \label{eq:1molMF} \\
		&& + t_{b1}  \sum_{\sigma}\hat  c^\dagger_{\mu 1\sigma} \hat  c_{\nu 2\sigma}  
		+ U \sum_i \hat n_{\mu i\uparrow} \hat n_{\mu i\downarrow}\notag\\
		&&	+ V_{b1} \hat n_{\mu1} \hat n_{\mu2}
		+ J_{b1} \hat{\bm S}_{\mu1} \cdot \hat{\bm S}_{\mu2}, \notag
\end{eqnarray}
where we have used the fact that $\bm{\lambda}_{b1}=\bm{0}$ in all materials studied (Table \ref{tab:ts}). 
This can be solved exactly and has two degenerate ground states, which describe the spin-one--half degrees of freedom on the dimer in the  insulating phase. However, the spin does not, in general, have equal weights on the two molecules within the dimer, the charge disproportionation is given by
\begin{eqnarray}
	\langle \hat{n}_{\mu1} - \hat{n}_{\mu2}\rangle = \frac{\Delta\left( \Delta + \sqrt{\Delta^2+t_{b1}^2} \right)}{t_{b1}^2 + \Delta\left( \Delta + \sqrt{\Delta^2+t_{b1}^2} \right)}.
\end{eqnarray}

\begin{figure}
	\begin{center}
		\includegraphics[width=0.9\columnwidth]{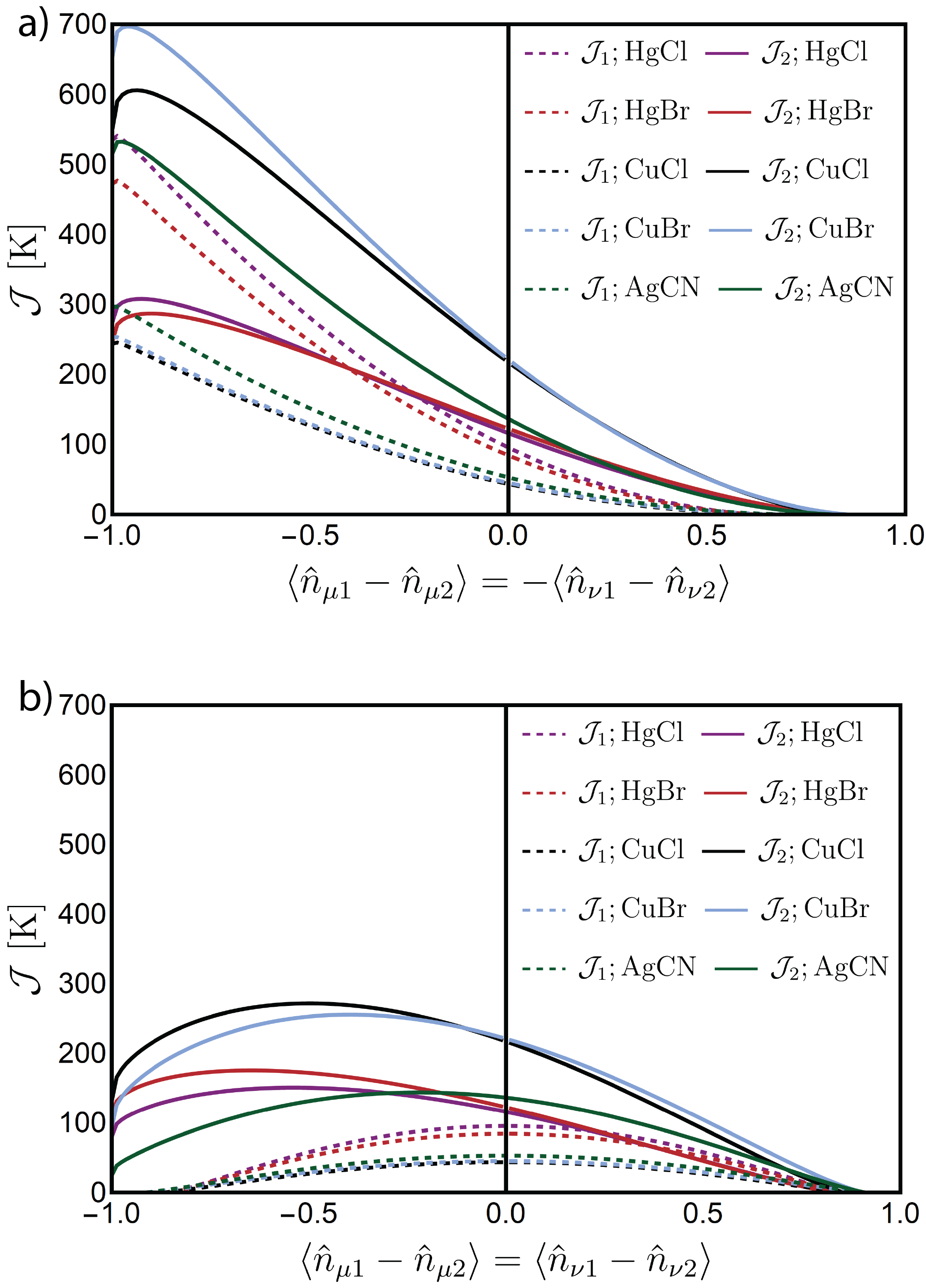} 
		\caption{Variation of the effective interdimer superexchange interaction with the charge disproportionation, $\langle \hat{n}_{\mu1}-\hat{n}_{\mu2}\rangle$, for (a) antiferrodipolar ordering and (b) ferrodipolar ordering. For ease of comparison both panels use the same axis scales.}\label{fig:J_vs_Deltan} 
	\end{center}
\end{figure}

Scalar hopping between dimers of the same  sublattices, Eq. (\ref{eq:tb2}),  induces an effective spin-spin symmetric exchange interaction, ${\mathcal J}_1$, between the electrons localized on the dimers at second order. Similarly one finds a Heisenberg interaction, ${\mathcal J}_2$, between spins on different sublattices caused by virtual hopping between neighboring dimers,  Eq.  (\ref{eq:tpq}). We calculate this via the methods described in \cite{powell17prb,powell17prl} and plot the dependence of ${\mathcal J}_{n}$ on the dipole strength, $\langle \hat{n}_{\mu1} - \hat{n}_{\mu2} \rangle$ in Fig. \ref{fig:J_vs_Deltan}. Here we consider two cases: panel (a)  shows the results when charge disproportionation takes equal magnitude but opposite sign on the two sites ($\langle \hat{n}_{\mu1}-\hat{n}_{\mu2}\rangle=-\langle \hat{n}_{\nu1}-\hat{n}_{\nu2}\rangle$), \textit{i.e.}, an antiferrodipolar arrangement; and panel (b) considers a ferrodipolar arrangement where $\langle \hat{n}_{\mu1}-\hat{n}_{\mu2}\rangle=\langle \hat{n}_{\nu1}-\hat{n}_{\nu2}\rangle$.

\begin{figure}
\begin{center}
	\includegraphics[width=0.8\columnwidth]{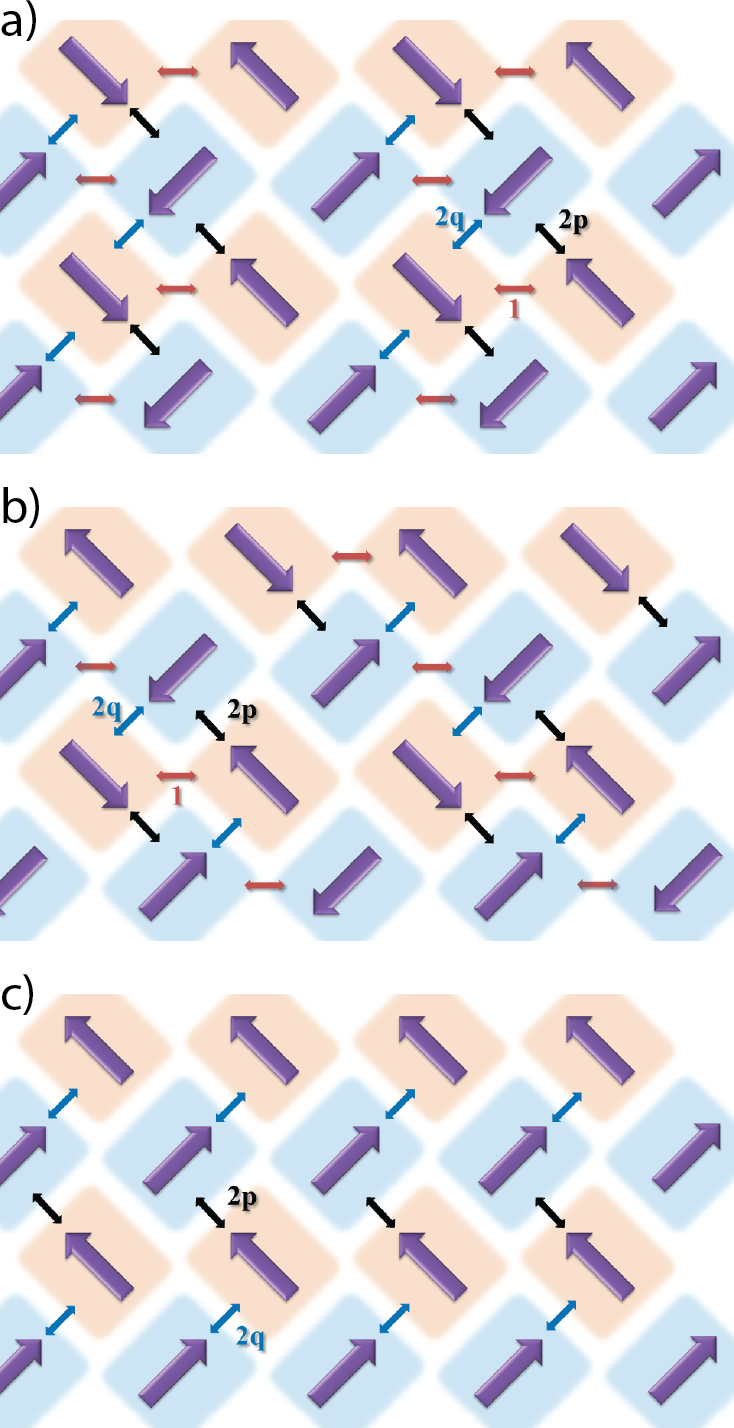}
	\caption{Possible dipole ordering patterns (a,b) in the quasi-one-dimensional limit of the TFIM ($K_1\gg K_2$) and (c) in the square lattice limit of the TFIM, ($K_2\gg K_1$). The head of the arrows indicates the monomer that contains fewer electrons than the monomer at the tail of the arrow.  To understand the microscopic dipolar order shown here it is helpfull to compare with Fig. \ref{fig:lattice}. Only the  supperexchange interactions that are strongest when the dipoles order are marked in this figure. In all cases strong charge disproportionation combined with  dipole order leads to a quasi-one-dimensional spin system.}\label{fig:dipole_order} 
\end{center}
\end{figure}

\begin{figure*}
\begin{center}
	\includegraphics[width=2\columnwidth]{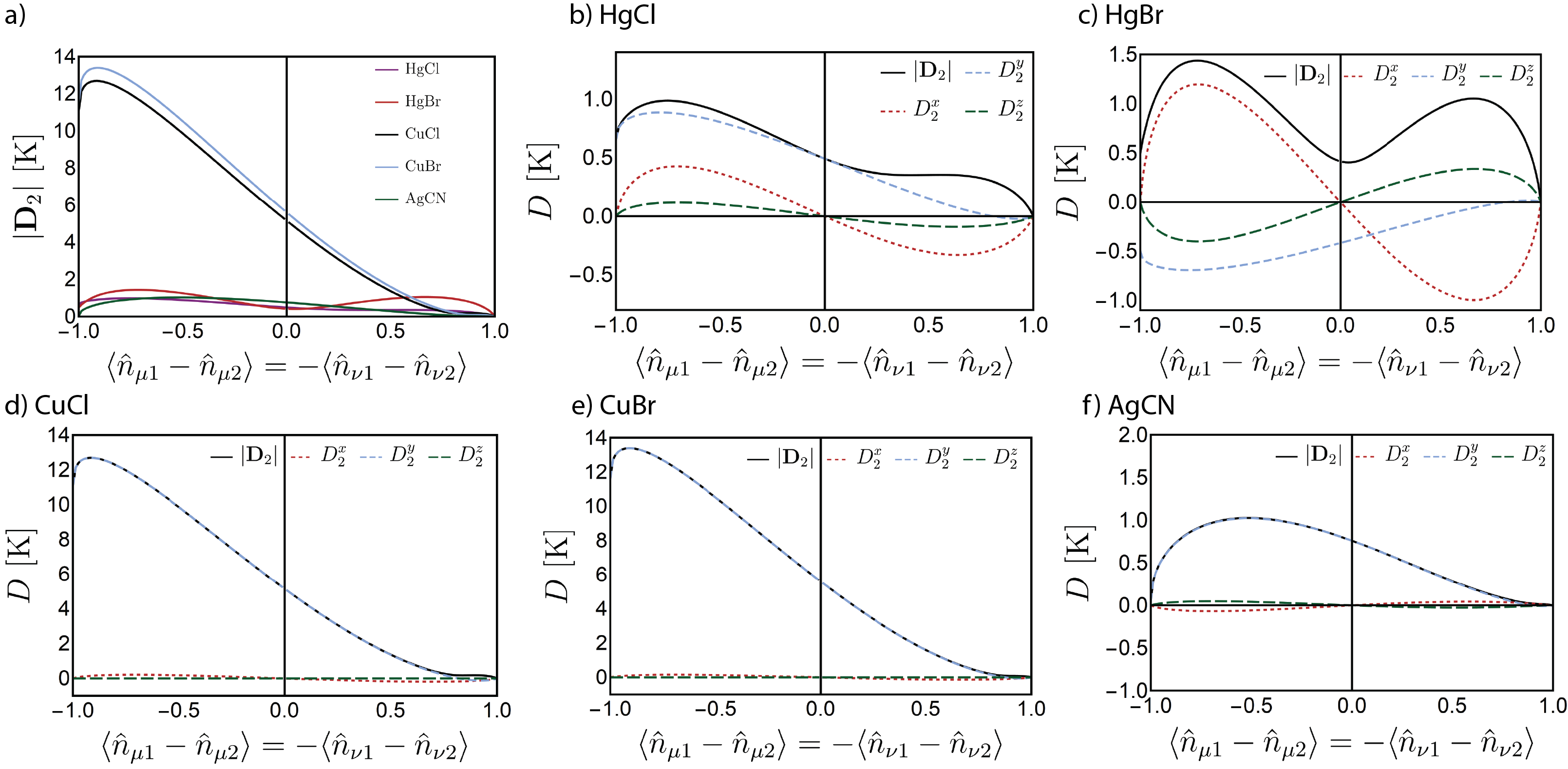}
	\caption{Variation of the effective interdimer Dzyaloshinskii-Moriya antisymmetric exchange with the charge disproportionation, $\hat{n}_{\mu1}-\hat{n}_{\mu2}$, for an antiferrodipolar configuration of dimers. a) Magnitude for all materials and Cartesian components for b) \HgCl, c)  \HgBr, d)  \CuCl, e)  \CuBr, and f)  \AgCN. }\label{fig:D_vs_DeltanAFD} 
\end{center}
\end{figure*}

\begin{figure*}
\begin{center}
	\includegraphics[width=2\columnwidth]{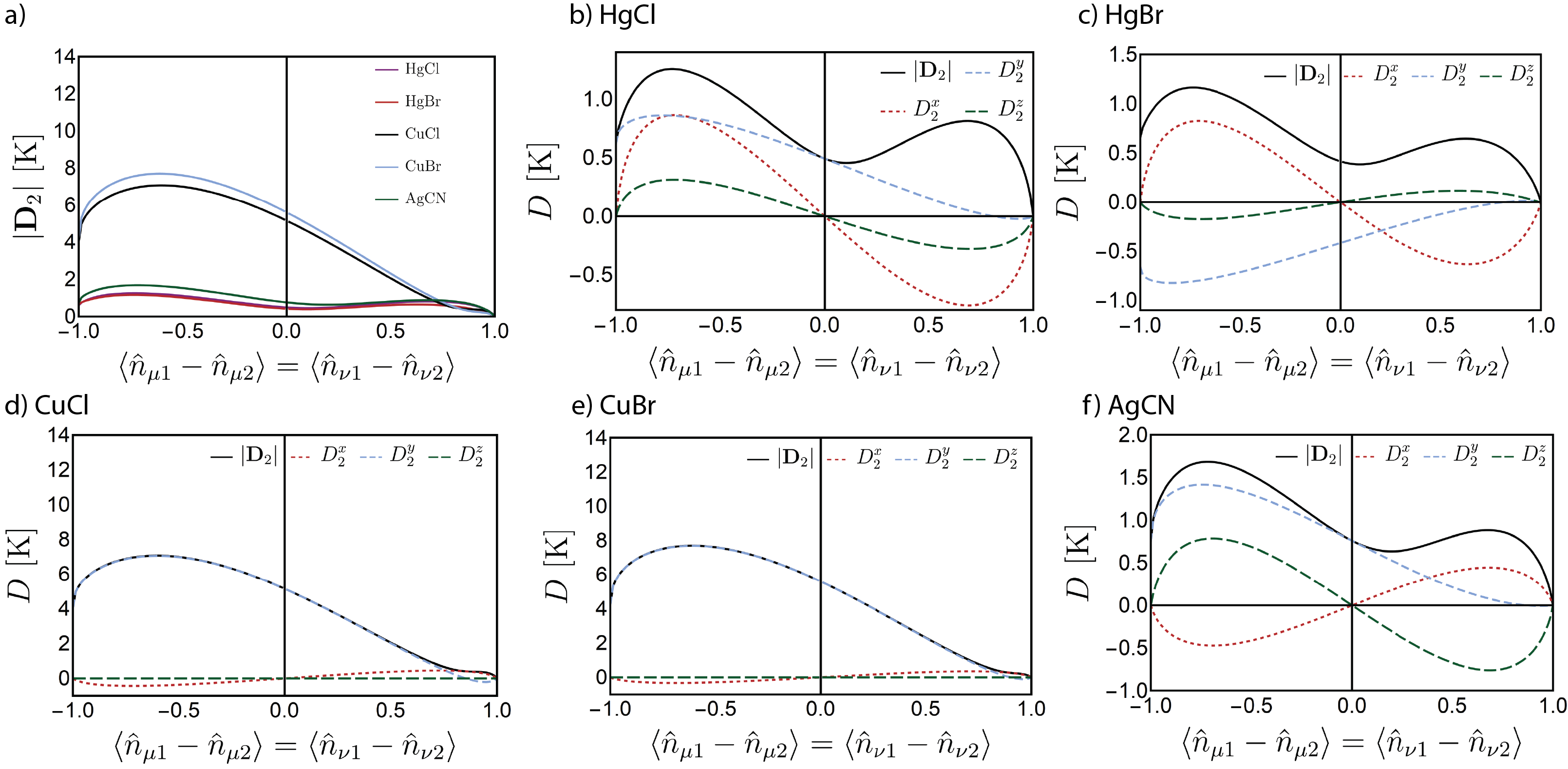} 
	\caption{Variation of the effective interdimer Dzyaloshinskii-Moriya antisymmetric exchange with the charge disproportionation, $\hat{n}_{\mu1}-\hat{n}_{\mu2}$, for a ferrodipolar configuration of dimers. a) Magnitude for all materials and Cartesian components for b) \HgCl, c)  \HgBr, d)  \CuCl, e)  \CuBr, and f)  \AgCN. }\label{fig:D_vs_DeltanFD} 
\end{center}
\end{figure*}

In both cases the variation of the interdimer superexchange with charge disproportionation is extremely dramatic. This is a simple consequence of the dependence of the perturbative processes, in the monomer model (Fig. \ref{fig:whyXTT}), on the occupation of the two different Wanniers, which is  easily understood by considering the limits $\langle \hat{n}_{\mu1}-\hat{n}_{\mu2}\rangle\rightarrow\pm1$.

Notice that  superexchange interactions vanish for $\langle \hat{n}_{\mu1}-\hat{n}_{\mu2}\rangle\rightarrow1$ in both the antiferrodiplar and ferrodipolar cases. This can easily be understood by examining the form of the interdimer hopping, Eqs. (\ref{eq:tb2}) and (\ref{eq:tpq}), and  Fig. \ref{fig:whyXTT}. Clearly, all virtual hopping processes between the dimers cease when $\langle \hat{n}_{\mu1}-\hat{n}_{\mu2}\rangle\rightarrow1$ as they require a partial occupation of monomer $\mu2$. 

For $\langle \hat{n}_{\mu1}-\hat{n}_{\mu2}\rangle\rightarrow-1$ each of the cases gives a different result. ${\cal J}_1=0$ in the ferrodipolar case. This is again a straightforward consequence of the detailed form of the interdimer hopping, Eq. (\ref{eq:tb2}) see also Fig. \ref{fig:whyXTT}, which shows that superexchange via the $t_{b2}$ pathway requires partial occupation of Wannier $\nu1$, which is doubly occupied in this limit. In the antiferrodipolar case ${\cal J}_1\not\rightarrow0$ for $\langle \hat{n}_{\mu1}-\hat{n}_{\mu2}\rangle\rightarrow-1$. In this limit $\langle \hat{n}_{\mu2}\rangle=\langle \hat{n}_{\nu1}\rangle\rightarrow1$ and the second order perturbation due to $t_{b2}$, Eq. (\ref{eq:tb2}), reduces to the usual case of superexchange between two half-filled sites \cite{powellNotes}. Similarly  ${\cal J}_2$ simplifies to the problem of superexchange between two half-filled sites when $\langle \hat{n}_{\mu1}-\hat{n}_{\mu2}\rangle\rightarrow-1$. For an antiferrodipolar arrangement ${\cal J}_2\rightarrow {\cal J}_{2p}\propto t_p^2$ and for a ferrodipolar arrangement ${\cal J}_2\rightarrow {\cal J}_{2q}\propto t_q^2$ and hence ${\cal J}_{2p}>{\cal J}_{2q}$. 

To understand the impact of dipole order on the magnetic properties it is important to know which long-range pattern of dipoles is realized. The  solution of the TFIM, Eq. (\ref{eq:HIsing}), is an extremely challenging problem and beyond the scope of this Article. But, for illustrative purposes it is useful to consider the order in the classical zero-field regime. Two low-energy states for $K_1\gg K_2$ are sketched in Figs. \ref{fig:dipole_order}a and b, and the lowest energy   state for $K_2\gg K_1$ is sketched in Fig. \ref{fig:dipole_order}c. In all cases we have marked only the strongest magnetic superexchange interactions. The magnetic interactions that are not shown are zero for complete charge disproportionation $|\langle \hat{n}_{\mu1}-\hat{n}_{\mu2}\rangle|\rightarrow1$, but simply small in the more realistic case of incomplete charge disproportionation $0<|\langle \hat{n}_{\mu1}-\hat{n}_{\mu2}\rangle|<1$. The values of the `large' and `small' interactions correspond to changing the sign of charge disproportionation $\langle \hat{n}_{\mu1}-\hat{n}_{\mu2}\rangle$ in Fig. \ref{fig:J_vs_Deltan}.

The order sketched in Figs. \ref{fig:dipole_order}a and b are relevant to \HgCl and \HgBr. Our calculations suggest that these are very close in energy, with the very weak $W^{TT}_3$ interaction, cf. Table \ref{tab:Ising}, favoring the order sketched in \ref{fig:dipole_order}b. However, the existence of competing orders very interesting given the recent observation of dipole liquid-like behaviors in \HgBr \cite{DrichkoScience}. This suggests the possibility of resonance structures similar to those proposed in spin liquids \cite{powellScience,andersonRVB}.

In the dipole orders sketched in Fig. \ref{fig:dipole_order} the `strong' bonds form quasi-one-dimensional spin systems, coupled by the `weak' bonds. The effects of this anisotropy will depend strongly on the difference between the $\cal J$ values of the `strong' and `weak' bonds, which is in turn determined by the magnitude of the charge disproportionation. The resulting spin model could range from a subtly anisotropic quasi-two-dimensional system (for weak charge disproportionation) to quasi-one-dimensional spin systems (for large charge disproportionation). However, we note that the geometrical frustration of the magnetic interactions means that this quasi-one-dimensional limit is more stable than one might naively expect \cite{powell17prl,kenny,kennyACIE}. Therefore, the effective one-dimensionalization of the system when it is dipole ordered could be important for understand the large ``dipole-solid+spin-liquid'' phase found by Hotta in her exact diagonalization studies of the NIH model \cite{Hotta}.

Similarly, the vector hopping between the different sublattices (cf. Table \ref{tab:ts}) induces a Dzyaloshinskii-Moriya  antisymmetric exchange interaction:
\begin{eqnarray}
\hat{H}_\text{DM} = \sum_{\langle\mu,\nu\rangle} {\bm D}_{1} \cdot \hat{\bm S}_{\mu} \times \hat{\bm S}_{\nu}
+  \sum_{[\mu,\nu]} {\bm D}_{2} \cdot \hat{\bm S}_{\mu} \times \hat{\bm S}_{\nu}.
\end{eqnarray}
This is calculated as described in  \cite{powell17prb} and plotted in Fig. \ref{fig:D_vs_DeltanAFD}, for an antiferrodipolar arrangement, and Fig. \ref{fig:D_vs_DeltanFD}, for an ferrodipolar arrangement. ${\bm D}_1=\bm 0$ in all materials because there is an inversion symmetry between the dimers on the same sublattice. The Dzyaloshinskii-Moriya interaction is likely to be profoundly important for determining the macroscopic magnetic properties because of the low dimensionality of the system \cite{merminwagner}.

Regardless of the degree of charge order, the Dzyaloshinskii-Moriya interaction is much stronger in \CuBr and \CuCl than it is in \AgCN. This is interesting because the former pair both order antiferromagnetically, whereas the  long-range magnetic order has not been observed in \AgCN at any temperature. The Mermin-Wagner theorem  \cite{merminwagner} shows that there cannot be long-range order in two-dimensional Heisenberg model, but a Dzyaloshinskii-Moriya interaction violates the assumptions required to prove the theorem and  allows for long-range magnetic order. Therefore, a natural perspective on these materials, given the very week interlayer interactions, is that the absence of long-range order  is the natural state of affairs; and that a significant Dzyaloshinskii-Moriya interaction is required to cause long-range order. 

Finally we note that, in several compounds, dipole ordering significantly increases the  Dzyaloshinskii-Moriya coupling. Thus, dipole order could cause or enhance long-range antiferromagnetic order.

\section{Conclusions}

In summary, we have used a combination of first principles calculations and empirical  relationships for the Coulomb interactions to parameterize the  HIN model of coupled dipolar and spin degrees of freedom for \HgCl, \HgBr, \CuCl, \CuBr, and \AgCN. In all materials the coupling constants for the TFIM of the dipoles are the largest energy scales in the problem. This quantifies, justifies and confirms recent claims that the dipoles are of crucial importance for understanding these materials \cite{Clay2019,Hotta,Naka,Lunkenheimer,Abdel-Jawad,Sasaki}.

Two  effects are responsible for different ferroelectric behaviors of the \X salts. (i) The inter-dimer hopping, $t_{b1}$, which gives rise to the ``transverse field" in the Ising model for the dipoles ($H^T=2t_{b1}$), is between a third and a tenth smaller in  \HgCl and \HgBr than in \CuCl, \CuBr, and \AgCN. (ii) The TFIM of dipoles is in the quasi-one-dimensional limit for \HgCl and \HgBr, but quasi-two-dimensional (between the square and isotropic triangular limits) for \CuCl, \CuBr, and \AgCN. Thus, the dimensionless critical field is 3\,--\,5 times smaller in the former pair of compounds. Clearly then, effect (ii) is much more important for this conclusion than  effect (i).  This occurs because 
$X_{TT}\propto t_{b2}^2$ [see Eq. (\ref{eq:XTT})] and $t_{b2}$ is 25-50\% larger in \HgCl and \HgBr than the other materials, see Table \ref{tab:ts}, which means that $X_{TT}$ is much bigger in the mercuric materials. 
$Y_{TT}\propto (t_p^2 - t_q^2)$ [see Eq. (\ref{eq:YTT})] and $t_p$ is $\sim50\%$ larger in the non-mercuric  materials, which results in a much larger $Y_{TT}$ in these systems.

We have also found that dipolar order and even short-range dipolar correlations have a profound impact on the nature of the interdimer magnetic (superexchange) interactions. In particular for dipole crystals this appears to   drive the materials towards quasi-one-dimensional magnetic interactions, which could be important for understand the spin liquids found in many of these materials. Conversely, dipole order can increase with Dzyaloshinskii-Moriya interaction, which may tend to stabilize magnetic order. Understanding the interplay of these effects is a significant challenge for future quantum many-body theories.

\section*{Acknowledgments}
This work was supported by the Australian Research Council through grants DP160100060 and DP180101483. 

\appendix

\section{Parameters in Eqs. (\ref{eq:X}) and (\ref{eq:Y})}

The simplest method to calculate the contributions to the effective spin-dipole Hamiltonian at second order in $\hat{H}_1$ and $\hat{H}_2$ is to make use of projectors. Recall that the spin operator 
\begin{eqnarray}
\hat{P}_0 \equiv \frac14-\bm{S}_1\cdot\bm{S}_2=\left\{
\begin{array}{ll}
	1  & \text{if the spins form a singlet} \\ 
	0  & \text{if the spins form a triplet,}	
\end{array}\right.
\end{eqnarray} 
and
\begin{eqnarray}
\hat{P}_1 \equiv \bm{S}_1\cdot\bm{S}_2+\frac34=\left\{
\begin{array}{ll}
0  & \text{if the spins form a singlet} \\ 
1  & \text{if the spins form a triplet,}	
\end{array}\right.\,
\end{eqnarray} 
Similarly, $\hat{P}_\pm \equiv T^z \pm \frac12$  projects the unpaired electron on to monomers 1 ($-$) and 2 ($+$). These four projectors and the operators $T^\pm=T^x \pm iT^y$ make it straightforward, but tedious, to enumerate all possible second order corrections in the scalar hopping parameters, Eqs. (\ref{eq:tb2}) and  (\ref{eq:tpq}), and rearrange into the form of the effective Hamiltonains, Eqs. (\ref{eq:X}) and (\ref{eq:Y}). Whence, one finds that
	\begin{eqnarray}
	X_{TT} &=& \frac{t_{b2}^2}{U}
	+ \frac{t_{b2}^2\left( 2V_{b1} +7 J_{b1} \right)}{V_{b1}\left( V_{b1} - 2J_{b1} \right)}, \label{eq:XTT}
	\end{eqnarray}
	\begin{eqnarray}
	X_{T} &=& -\frac{t_{b2}^2}{2U}, \\
	X_{SSTT} &=& \frac{4t_{b2}^2}{U}
	-\frac{4t_{b2}^2 J_{b1}}{V_{b1}\left( V_{b1} - 2J_{b1} \right)}, \\
	X_{SST} &=& -\frac{2t_{b2}^2}{U}, \\
	X_{SS} &=& \frac{t_{b2}^2 J_{b1}}{V_{b1}\left( V_{b1} - 2J_{b1} \right)}, \\
	Y_{TT} &=& \frac{(t_{p}^2 - t_{q}^2)}{U}
	+ \frac{(t_{p}^2-t_{q}^2)(2V_{b1} + 7J_{b1})}{V_{b1}\left( V_{b1} - 2J_{b1} \right)}, \label{eq:YTT}\\
	Y_{zx} &=& -\frac{t_p t_q}{2U} -\frac{t_pt_q(2V_{b1} + 7J_{b1})}{2V_{b1}\left( V_{b1} - 2J_{b1} \right)}, \\
	Y_{T-} &=& -\frac{t_{p}^2}{2U}, \\
	Y_{T+} &=& -\frac{t_{q}^2}{2U}, \\
	Y_{x} &=& -\frac{t_p t_q}{4U} -\frac{t_pt_q(2V_{b1} + 7J_{b1})}{4V_{b1}\left( V_{b1} - 2J_{b1} \right)} \\
	Y_{SSTT} &=& \frac{4(t_{p}^2 - t_{q}^2)}{U}
	-\frac{4(t_{p}^2-t_{q}^2)J_{b1}}{V_{b1}\left( V_{b1} - 2J_{b1} \right)}, \\
	Y_{SSzx} &=&  \frac{2t_p t_q}{U} 
	-\frac{ t_pt_q (V_{b1} +J_{b1})}{V_{b1}\left( V_{b1} - 2J_{b1} \right)} \\
	Y_{SST-} &=& -\frac{2t_{p}^2}{U}, \\
	Y_{SST+} &=& -\frac{2t_{q}^2}{U}, \\
	Y_{SSx} &=& \frac{t_p t_q}{U}
	-\frac{t_pt_q(V_{b1} +J_{b1})}{V_{b1}\left( V_{b1} - 2J_{b1} \right)}, \\
	Y_{SS} &=& \frac{(t_{p}^2+t_{q}^2)J_{b1}}{V_{b1}\left( V_{b1} - 2J_{b1} \right)}, \\
	Y_{4Szx} &=& \frac{8 t_pt_qJ_{b1}}{V_{b1}\left( V_{b1} - 2J_{b1} \right)}, 
	\end{eqnarray}
	and
	\begin{eqnarray}
	Y_{4Sx} &=&  \frac{4 t_pt_qJ_{b1}}{V_{b1}\left( V_{b1} - 2J_{b1} \right)}.
	\end{eqnarray}


\bibliography{refs}

\end{document}